\documentclass{iau} 
\usepackage{graphicx}

\title[] 
{Learning about AGB stars by studying the stars polluted by their outflows}

\author[Escorza \& De Rosa]   
{Ana Escorza$^1$ \and
Robert J. De Rosa$^1$
}

\affiliation{$^1$European Southern Observatory, Alonso de C\'{o}rdova 3107, Vitacura, Santiago, Chile \\ email: {\tt ana.escorza@eso.org} \\[\affilskip]
}

\pubyear{2022}
\volume{366}  
\jname{The origin of outflows in evolved stars}
\editors{L. Decin, A. Zijlstra \& C. Gielen, eds}
\begin{document}

\maketitle

\begin{abstract}
A rich zoo of peculiar objects forms when Asymptotic Giant Branch (AGB) stars, undergo interactions in a binary system. For example, Barium (Ba) stars are main-sequence and red-giant stars that accreted mass from the outflows of a former AGB companion, which is now a dim white dwarf (WD). Their orbital properties can help us constrain AGB binary interaction mechanisms and their chemical abundances are a tracer of the nucleosynthesis processes that took place inside the former AGB star. The observational constraints concerning the orbital and stellar properties of Ba stars have increased in the past years, but important uncertainties remained concerning their WD companions. In this contribution, we used HD\,76225 to demonstrate that by combining radial-velocity data with Hipparcos and Gaia astrometry, one can accurately constrain the orbital inclinations of these systems and obtain the absolute masses of these WDs, getting direct information about their AGB progenitors via initial-final mass relationships.

\end{abstract}

\firstsection 
\section{Introduction}

Asymptotic Giant Branch (AGB) stars are the main producers of about half of the chemical elements heavier than iron via the slow neutron capture (s-) process of nucleosynthesis \cite[(e.g. Lugaro et al. 2003; Karakas 2010; K\"{a}ppeler et al. 2011)]{Lugaro03, Karakas10, Kappeler11}. However, the determination of individual atomic abundances on the surface of AGB stars is complicated by their complex convective envelopes and circumstellar environments, their high mass-loss rates, and the presence of broad molecular bands on their spectra \cite[(e.g. Busso et al. 2001; Van Eck et al. 2017; Shetye et al. 2018)]{Busso01, VanEck17, Shetye18}. Additionally, a significant fraction of AGB stars might have stellar or sub-stellar companions, which are difficult to detect and characterise also due to the previously mentioned factors \cite[(e.g. Mayer et al. 2014; Decin et al. 2020)]{Mayer14, Decin20}. With this contribution, we want to convince the reader that we can use AGB binary interaction products to learn about the AGB phase, specifically about binary interaction and nucleosynthesis processes.

Barium (Ba) stars are a prototypical example of AGB binary interaction products. These chemically peculiar stars formed when an AGB star transferred mass to its unevolved companion in a binary system \cite[(e.g. McClure 1984; Udry et al. 1998; Jorissen et al. 1998)]{McClure84, Udry98, Jorissen98}. The former AGB star evolved long ago and is now a dim white dwarf (WD) and its s-process enriched companion, the Ba star, is now the most luminous star in the system and can be observed at different evolutionary phases \cite[(e.g. Escorza et al. 2017, 2019; Jorissen et al. 2019, Shetye et al. 2020)]{Escorza17, Escorza19, Jorissen19, Shetye20}.

The fact that Ba stars are known products of binary interaction means that their orbital properties can provide constraints to binary interaction and evolution models \cite[(e.g. Bona{\v c}i{\'c} Marinovi{\'c} et al. 2008; Dermine et al. 2013; Escorza et al. 2020)]{BonacicMarinovic08, Dermine13, Escorza20}. Additionally, the surface s-process abundances of Ba stars are a tracer of the chemical production inside the AGB star that polluted them \cite[(e.g. De Castro et al. 2016, Karinkuzhi et al. 2018, Roriz et al. 2021)]{deCastro16, Karinkuzhi18, Roriz21}. Keeping mixing and dilution processes \cite[(e.g. Charbonnel et al. 2007, Stancliffe et al. 2007, Aoki et al. 2008)]{Charbonnel07, Stancliffe07, Aoki08} in mind, one can combine the properties of the two stellar components and the Ba star abundances to learn about nucleosynthesis models as well \cite[(e.g. Cseh et al. 2022)]{Cseh22}. When one tries to use Ba star observations to constrain models, the largest observational uncertainty comes from the WD companions since they are cool, dim, and directly undetectable in most systems. Their masses are a key input parameter to both binary and nucleosynthesis models, but very few absolute masses have been determined since these are single-lined spectroscopic systems, and there is normally no information about the orbital inclinations (\cite[Pourbaix \& Jorissen 2000]{Pourbaix00}, for example, published a few exceptions based on Hipparcos astrometry). \cite[Escorza \& De Rosa (in prep)]{Escorza22} combined radial-velocity (RV) data with astrometric measurements from the Hipparcos and Gaia missions to determine the astrometric orbital parameters and the companion masses of the Ba stars studied by \cite[Jorissen et al. (2019)]{Jorissen19} and \cite[Escorza et al. (2019)]{Escorza19}. This short contribution presents a proof-of-concept of this methodology using HD\,76225.

\renewcommand{\arraystretch}{1.5}
\begin{table}
\begin{center}
\caption{Overview of the main stellar properties of the HD\,76225 \cite[(Escorza et al. 2019)]{Escorza19}}\label{table1}
\vspace{1mm}
\begin{tabular}{|c|c|c|c|c|c|}
\hline
{\bf T$_{eff}$ [K]} & {\bf logg [dex]} & {\bf [Fe/H]} & {\bf [s/Fe]$^{(1)}$} & {\bf L/L$_{\odot}$} & {\bf M$_1$/M$_{\odot}$} \\
6340 $\pm$ 50 & 3.9 $\pm$ 0.2 & -0.37 $\pm$ 0.08 & 1.25 $\pm$ 0.08 & 5.9 $\pm$ 0.7 & 1.21 $\pm$ 0.06\\
\hline
\end{tabular}
\end{center}
\vspace{1mm}
{\footnotesize
{\bf $^{(1)}$} [s/Fe] from Allen \& Barbuy (2006b).}
\end{table}

\renewcommand{\arraystretch}{1.5}
\begin{table}
\begin{center}
\caption{Overview of the main orbital properties of HD\,76225 \cite[(Escorza et al. 2019)]{Escorza19}}\label{table2}
\vspace{1mm}
\begin{tabular}{|c|c|c|c|c|c|c|}
\hline
{\bf P [days]} & {\bf ecc} & {\bf T$_0$ [HJD]} & {\bf $\omega$ [$^{\circ}$]} & {\bf K$_1$ [km\,s$^{-1}$]} & {\bf $\gamma$ [km\,s$^{-1}$]} & {\bf f(m) [M$_{\odot}$]}\\
2410 $\pm$ 2 & 0.098 $\pm$ 0.005 & 2451159 $\pm$ 400 & 267 $\pm$ 3 & 6.11 $\pm$ 0.04 & 30.34 $\pm$ 0.02 & 0.0561 $\pm$ 0.0010\\
\hline
\end{tabular}
\end{center}
\end{table}

\section{Stellar and binary properties of HD\,76225}

\begin{figure}[t]
\includegraphics[width=0.49\textwidth]{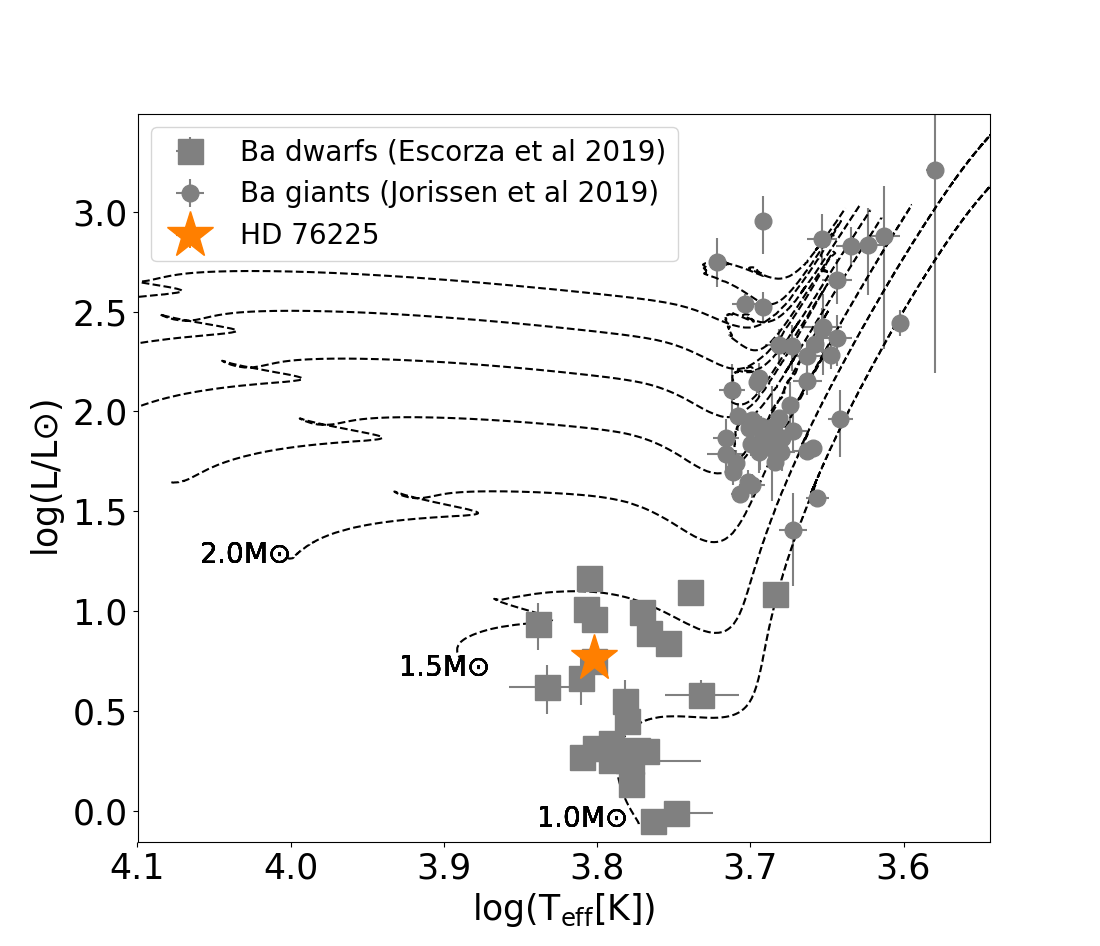} 
\includegraphics[width=0.49\textwidth]{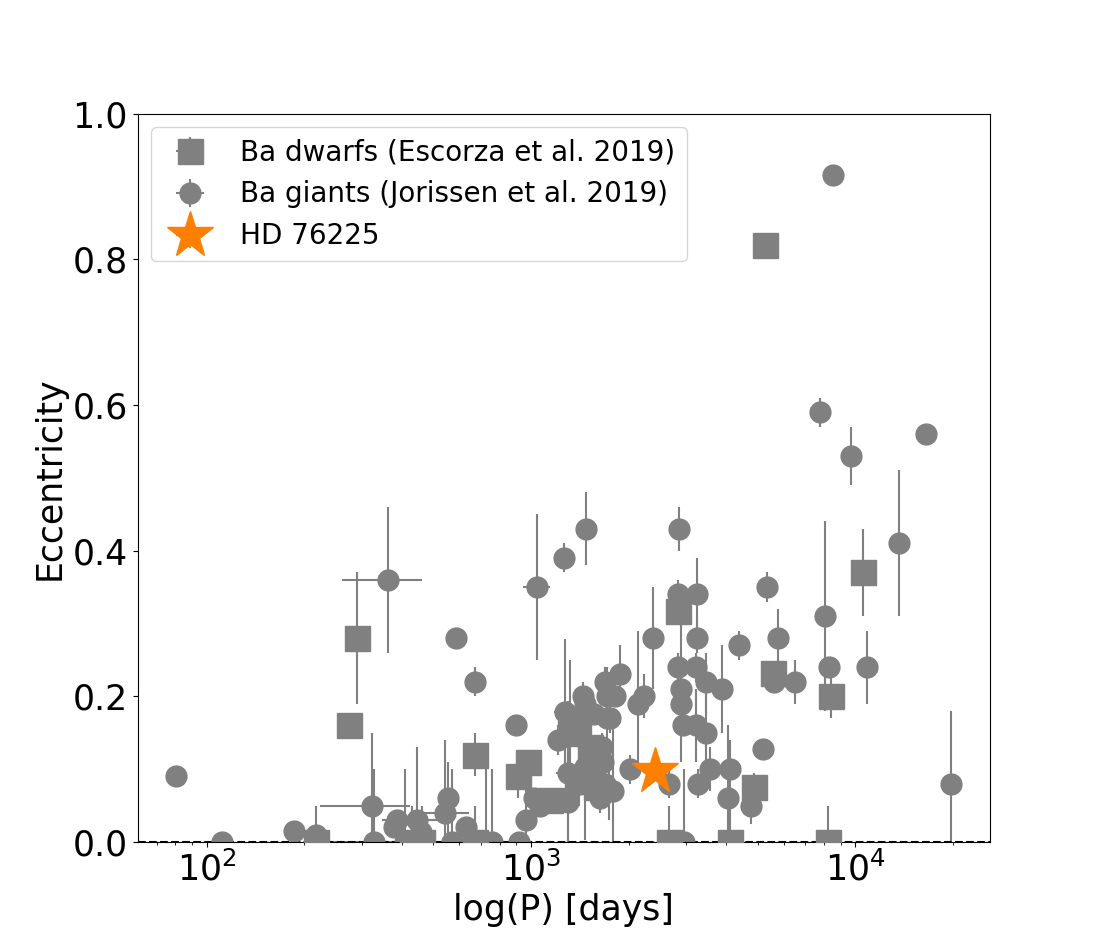} 
\caption{Hertzsprung-Russel diagram (left) and eccentricity-period diagram (right) of Ba dwarf and giants highlighting the location of the Ba dwarf HD\,76225.}\label{fig1}
\end{figure}

HD\,76225 (HIP\,43703) is a main-sequence Ba star, first proposed as such by \cite[North et al. (1994)]{North94}. \cite[Allen \& Barbuy (2006ab)]{AllenBarbuy06I,AllenBarbuy06II} determined its surface chemical abundances, including the s-process elements, and \cite[Escorza et al. (2019)]{Escorza19} determined the spectroscopic stellar parameters, the luminosity and the mass of the Ba star primary, as well as the spectroscopic orbital elements of the system. These characteristics have been summarised in Tables \ref{table1} and \ref{table2}. Additionally, Fig. \ref{fig1} highlights the location of HD\,76225 on the Hertzsprung-Russel diagram (left) and on the eccentricity-period diagram (right) together with other well-known members of the Ba star family.

As far as the authors are aware, there is no direct evidence of the presence of a WD companion in the system, i.e. no UV excess has been reported as it has been the case for other systems \cite[(e.g. Böhm-Vitense et al. 2000; Gray et al. 2011)]{Bohm-Vitense00, Gray11}. However, its high s-process enhancement suggests that HD\,76225 underwent mass transfer from an AGB star and the spectroscopic mass-function determined by \cite[Escorza et al. (2019)]{Escorza19} is compatible with the presence of a WD in the system (see Table \ref{table2}).

These observational constraints were obtained from spectra from the HERMES high-resolution spectrograph \cite[(Raskin et al 2011)]{Raskin11}, the CORAVEL spectrometer \cite[(Baranne et al. 1979)]{Baranne79}, and the FEROS spectrograph \cite[(Kaufer et al. 2000)]{FEROS}. However, in order to get the mass of the invisible companion in a single-lined spectroscopic binary, one needs the orbital inclination of the system and spectra alone is not enough to determine this.

\section{The determination of the WD companion's mass}

For this contribution, we combined the individual radial-velocity points from HERMES and CORAVEL published by \cite[Escorza et al. (2019)]{Escorza19} with astrometric data from Hipparcos and Gaia. We used the catalogue positions from the Hipparcos mission \cite[(Perryman et al. 1997)]{HippCat97} and the re-reduction of the intermediate astrometric data (IAD; \cite[van Leeuwen 2007]{vanLeeuwen07}). Additionally, since the latest data release of the Gaia mission, Early DR3 or EDR3 \cite[(Lindegren et al. 2021)]{GaiaEDR3}, does not contain individual astrometric measurements yet, we used the catalogue positions and proper motions and the Gaia predicted scan epochs and angles\footnote{Queried using the Gaia Observation Forecast Tool: https://gaia.esac.esa.int/gost/}. Following a similar methodology to that used by \cite[De Rosa et al. (2020), Kervella et al. (2020) and Venner et al. (2021)]{DeRosa20, Kervella20, Venner21}, among others, to determine the masses of exoplanets and using the code \textsc{orvara}, developed by \cite[Brandt et  al. (2021b)]{Brandt21}, we fit a single Keplerian model to all the different data sets at the same time employing a parallel-tempering Markov chain Monte Carlo (\textsc{ptmcmc}, \cite[Foreman-Mackey et al. 2013]{PTMCMC}).

\textsc{orvara} also uses the Hipparcos-Gaia Catalog of Accelerations (HGCA, \cite[Brandt 2018]{Brandt18}) to ensure the proper cross-calibration of the two astrometric data sets when comparing their proper motions. HD\,76225 shows a significant astrometric acceleration (proper motion difference between Hipparcos and Gaia) which is a key constraint to obtain an accurate and precise measurement of the two stellar masses. The code first fits the RV data, allowing RV points from each instrument to have a different RV zero point. Then the absolute astrometry is included and fit for the five astrometric parameters (positions, $\alpha$ and $\delta$, proper motions, $\mu_{\alpha}$ and $\mu_{\delta}$, and parallax, $\varpi$) using \textsc{htof} \cite[(Brandt et al. 2021a]{htof}) at each MCMC step. On top of the five astrometric parameters, we fit 10 parameters: the six Keplerian orbital elements (semimajor axis, $a$, eccentricity, $e$, time of periastron passage, T$_0$, argument of periastron, $\omega$, orbital inclination, $i$, and longitude of the ascending node, $\Omega$), the masses of the two components (M$_{Ba}$ and M$_{WD}$) and a RV jitter per instrument (one for CORAVEL and one for HERMES) to be added to the RV uncertainties. 

We assumed uninformative priors for all the orbital elements, but we adopted a Gaussian prior for the primary mass. We used the value given in Table \ref{table1} but used 3 times the error bar as sigma to account for systematic errors not accounted for in the statistical uncertainty. We used 15 temperatures and for each temperature we use 100 walkers with 100,000 steps per walker. The MCMC chains converged quite quickly, but we discarded the first 300 recorded steps (the first 15000 overall, as we saved every 50) as the burn-in phase to produce the results presented in Sect \ref{results}.

For more details about the computational implementation in \textsc{orvara} and \textsc{htof} and for case studies showing the performance of the code we refer to the mentioned publications.

\begin{figure}[t]
\centering
\includegraphics[width=0.8\textwidth]{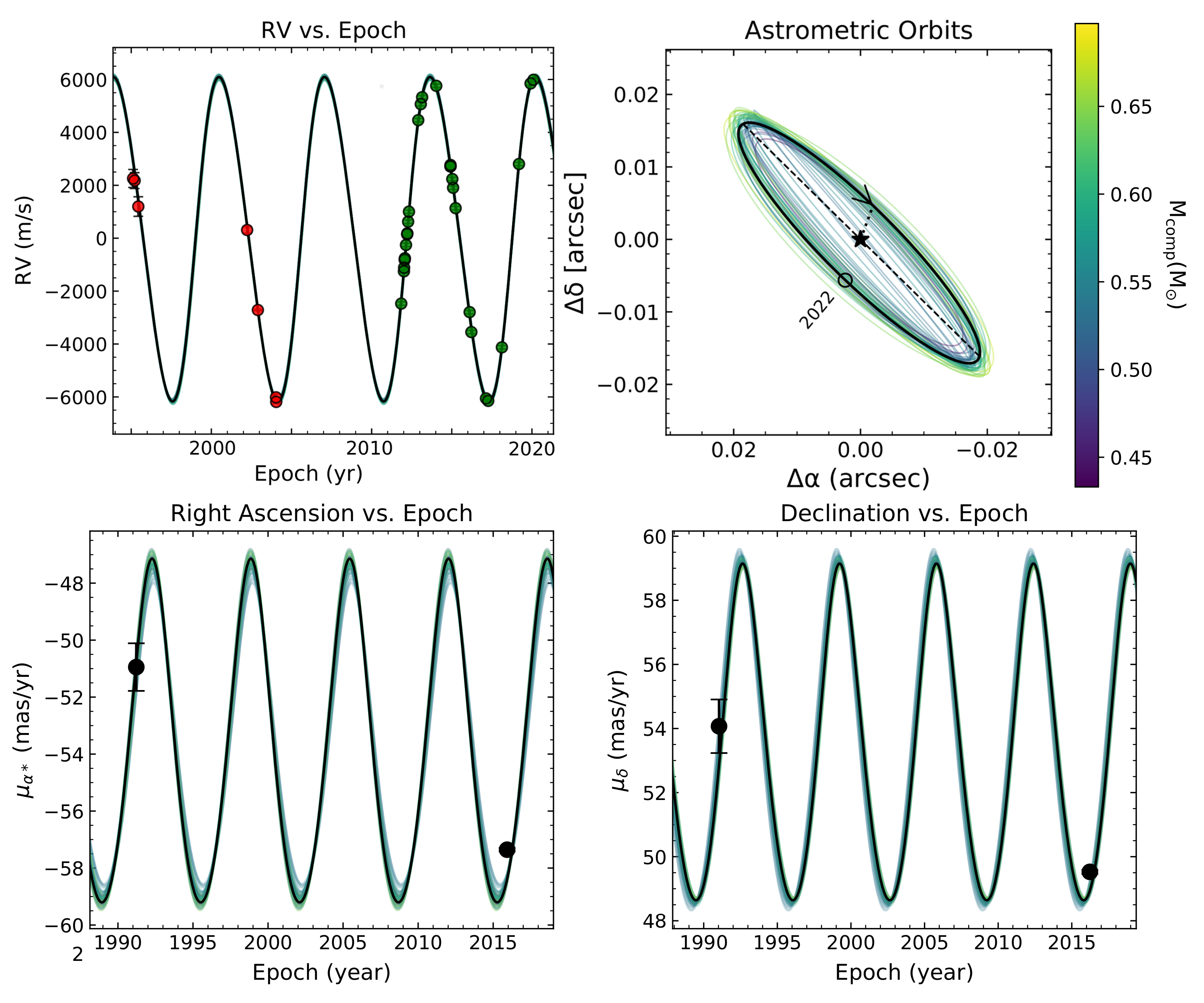} 
\caption{Fit to the RV data (top left, with CORAVEL data prior to 2005 and HERMES data newer than 2010) and to the Hipparcos (at the 1991.25 epoch) and Gaia (at the 2015.5 epoch) proper motions (bottom plots). The top right plot shows the projection of the orbit on the sky. The orbit with the highest likelihood is plotted as a thicker black line and 40 additional orbits are included, colour-coded as a function of the WD mass. }
\label{fig2}
\end{figure}

\section{Results and conclusions}\label{results}

\begin{figure}[t]
\centering
\includegraphics[width=0.94\textwidth]{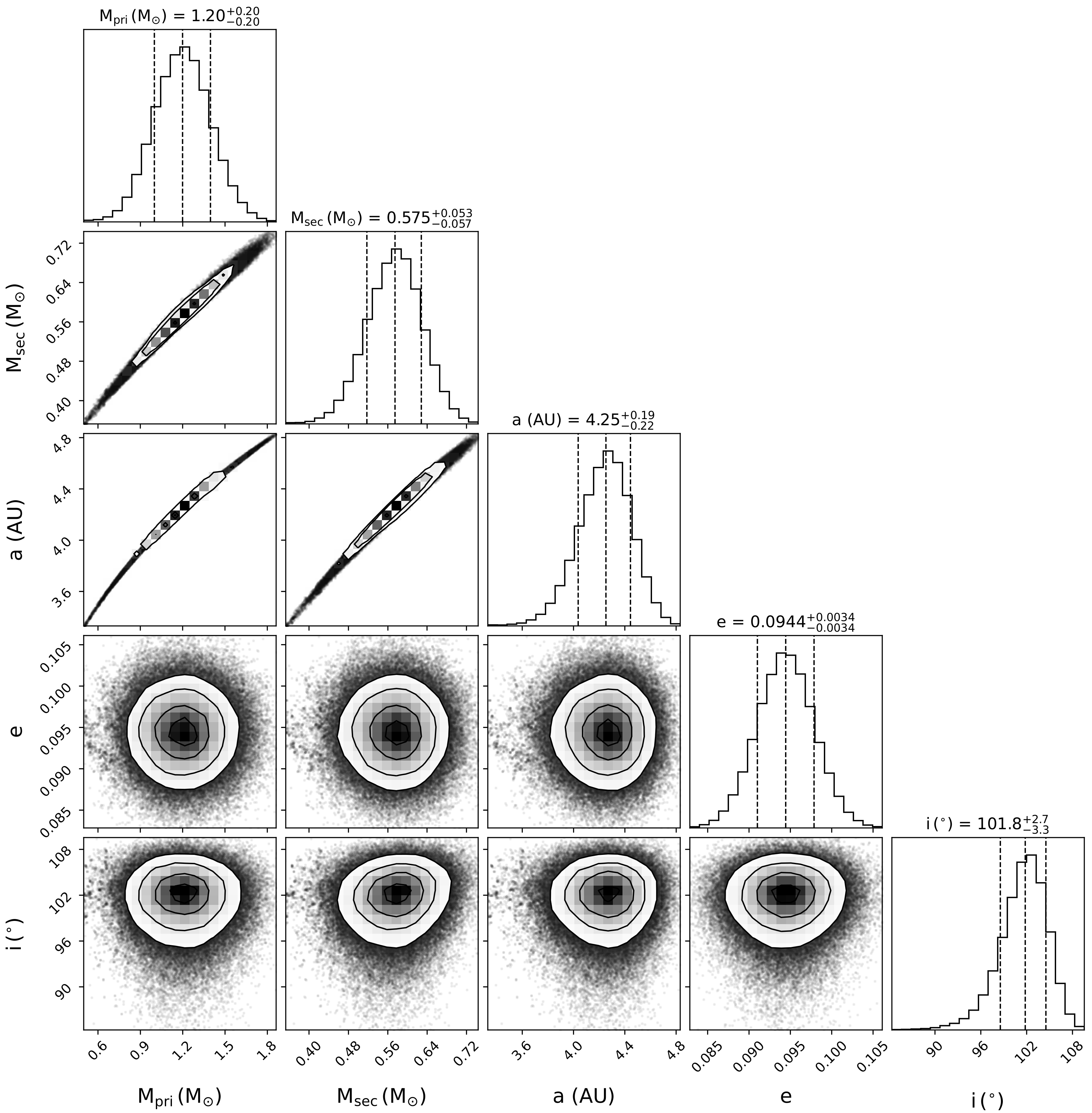} 
\caption{Corner plot of some of the derived parameters.}\label{fig3}
\end{figure}

Table \ref{table3} shows the derived posterior probabilities obtained from the MCMC orbital fit. Figure \ref{fig2} shows the fit to the RV data (top left) and to the absolute Hipparcos-Gaia astrometry (bottom plots) as well as the projection of the orbit on the sky (top right). In each plot, the orbit with the maximum-likelihood is plotted with a thicker black line, but we included 40 additional solutions for different WD masses. Finally, Fig. \ref{fig3} is a corner plot that shows the correlations among some of the derived parameters, especially the masses and the semimajor axis.

\renewcommand{\arraystretch}{1.2}
\begin{table}
\begin{center}
\caption{Overview of the MCMC results}\label{table3}
\vspace{1mm}
\begin{tabular}{l c|l c}
\hline
{\bf Parameter} & {\bf Median $\pm$ 1$\sigma$} & {\bf Parameter} & {\bf Median $\pm$ 1$\sigma$}\\
\hline
\textbf{Period, $P$ [days]} & 2405 $\pm$ 2 & \textbf{Parallax, $\varpi$ [mas]} & 5.91 $\pm$ 0.03 \\
\textbf{Eccentricity, $e$} & 0.094 $\pm$ 0.003 & \textbf{Ascending node, $\Omega$ [$^{\circ}$]} & 49 $\pm$ 2 \\
\textbf{Semimajor axis, $a$ [AU]} & 4.3 $\pm$ 0.2 & \textbf{Inclination [$^{\circ}$]} & 102 $\pm$ 3 \\
\textbf{Argument of periastron, $\omega$ [$^{\circ}$]} & 266 $\pm$ 2 & \textbf{Primary mass [$M_{\odot}$]} & 1.2 $\pm$ 0.2 \\
\textbf{Time of periastron, $T_{0}$ [HJD]} & 2455974 $\pm$ 11 & \textbf{Secondary mass [$M_{\odot}$]} & 0.58 $\pm$ 0.06 \\
\hline
\end{tabular}
\end{center}
\end{table}

Our joint orbital fit of the RV and astrometric data yields a period of 2405 $\pm$ 2 days and an eccentricity of 0.094 $\pm$ 0.003, which, in spite of the small uncertainties, are almost compatible with the results published by \cite[Escorza et al. (2019)]{Escorza19} using only RV data (see Table \ref{table2}). Additionally, we get a well constrained inclination for the system of 102$^{\circ}$ $\pm$ 3$^{\circ}$. This information combined with the primary mass, for which we had prior independent information, led to a secondary mass of 0.58 $\pm$ 0.06 $M_{\odot}$. Now, using an initial–final mass relationship (IFMR), we can estimate the initial mass of the AGB companion that polluted HD\,76225. Using the IFMR published by \cite[El-Badry et al. (2018)]{IFMR} from Gaia DR2, we obtained 1.8 $\pm$ 0.7 $M_{\odot}$. This mass is compatible with the idea that low-mass AGB stars ($<$\,3\,M$_{\odot}$; \cite[e.g. Lugaro et al. 2003; Karinkuzhi et al. 2018]{Lugaro03, Karinkuzhi18}) are responsible for the pollution of Ba stars. One could now use this information together with the abundance pattern on HD\,76225 and put constraints on mass transfer and dilution mechanisms \cite[(as done for example by Stancliffe 2021 and Cseh et al. 2022)]{Stancliffe21, Cseh22}. Finally, when the masses of the full sample are available, we will be able to look for correlations between the orbital parameters of Ba star systems and the WD masses and hopefully put constraints on mass-transfer mechanisms too.

\end{document}